\documentclass[11pt]{article}
\usepackage{amsmath,amssymb,color,graphics,epsfig}

\usepackage{amsfonts}

\newcommand{\be}{\begin{equation}}
\newcommand{\ee}{\end{equation}}
\newcommand{\bea}{\setlength\arraycolsep{2pt} \begin{eqnarray}}
\newcommand{\eea}{\end{eqnarray}}

\def\ft#1#2{{\textstyle{\frac{\scriptstyle #1}{\scriptstyle #2} } }}
\def\fft#1#2{{\frac{#1}{#2}}}

\def\0{{\sst{(0)}}}
\def\1{{\sst{(1)}}}
\def\2{{\sst{(2)}}}
\def\3{{\sst{(3)}}}
\def\4{{\sst{(4)}}}
\def\5{{\sst{(5)}}}
\def\6{{\sst{(6)}}}
\def\7{{\sst{(7)}}}
\def\8{{\sst{(8)}}}
\def\sst#1{{\scriptscriptstyle #1}}

\begin{document}

\begin{flushright}
\end{flushright}

\vspace{25pt}
\begin{center}
{\large {\bf Electrically-Charged Lifshitz Spacetimes, and Hyperscaling Violations}}

\vspace{10pt}
Zhong-Ying Fan and H. L\"u

\vspace{10pt}

{\it Department of Physics, Beijing Normal University, Beijing 100875, China}

\vspace{40pt}

\underline{ABSTRACT}
\end{center}

Electrically-charged Lifshitz spacetimes are hard to come by.  In this paper, we construct a class of such solutions in five dimensional Einstein gravity coupled to Maxwell and $SU(2)$ Yang-Mills fields.  The solutions are electrically-charged under the Maxwell field, whose equation is sourced by the Yang-Mills instanton(-like) configuration living in the hyperbolic four-space of the Lifshitz spacetime.  We then introduce a dilaton and construct charged and colored Lifshitz spacetimes with hyperscaling violations. We obtain a class of exact Lifshitz black holes.  We also perform similar constructions in four dimensions.

\vfill {\footnotesize Emails: zhyingfan@gmail.com \ \ \ mrhonglu@gmail.com}

\thispagestyle{empty}

\pagebreak



\newpage

\section{Introduction}

One way to break the Lorentz invariance in the AdS/CFT correspondence is to generalize the AdS (anti-de Sitter) spacetime to become the Lifshitz geometry
\cite{Kachru:2008yh}
\be
ds^2=\ell^2 \Big(r^{2z} dt^2 + \fft{dr^2}{r^2} + r^2 dx^i dx^i\Big)\,.\label{lifsmet}
\ee
The metric is invariant under the scaling
\be
r\rightarrow \lambda^{-1} r\,,\qquad t\rightarrow \lambda^z t\,,\qquad
x^i\rightarrow \lambda x^i\,.
\ee
The AdS/CFT correspondence is expected to continue to hold such that classical gravity in the bulk is dual to some strongly coupled condensed matter theory (CMT) living in $(t,x^i)$ that exhibits such scaling symmetry.  Although the Lifshitz metric (\ref{lifsmet}) is homogeneous,  it is not Einstein.  This leads to a question as what type of matter can support such a geometry.  Furthermore, for application in the AdS/CMT, one would like to construct the Lifshitz spacetimes with bulk fields that  are not too exotic from the condensed matter point of view.

If we focus on two-derivative gravities, a natural choice of matter is a vector field.  The Ansatz that respects the Lifshitz symmetry is $A=q r^z dt$, where $q$ is a constant. Unfortunately it will not satisfy the Maxwell equation without a source, namely $d{*F}=0$, where $F=dA$. On the other hand, a massive vector (Proca) field provides its own source, with the equation $d{*F}=m^2 {*A}$.  Indeed, a variety of Lifshitz spacetimes were constructed using the Proca fields \cite{Taylor:2008tg}.  In the original paper \cite{Kachru:2008yh}, a four-dimensional theory is considered in which the source is provided by a 3-form field strength, namely
\be
d{*F}=c\, H_3\,.
\ee
However, in four dimensions, one can Hodge dualize this $H_3=dB_2$ to a 1-form associated with an axion $\chi$.  In the Hodge dual description, the massless vector $A$ eats $\chi$ and becomes a massive Proca field.  Thus the construction really uses a Proca field in disguise and the physical degrees of freedom of $A$ and $B_2$ cannot be diagonalized.  It turns out that this is rather general whenever a vector is involved.  The string theory embedding of Lifshitz spacetimes is certainly possible [3-12]. Many of the examples are rather complicated and involve some unrealistic irrational exponents $z$.  The simplest examples may be some $z=2$ Lifshitz spacetimes that can be obtained from some AdS wave solutions by the Kaluza-Klein circle reduction.  It turns out, as shown explicitly in \cite{Chemissany:2011mb}, that the Kaluza-Klein vector in the effective Lagrangian indeed eats an axion scalar and becomes massive.

The prejudice against a Proca field is that it is more exotic than the Maxwell field and its dual operator in a boundary field theory has less interesting application in condensed matter systems.  We thus would like to construct Lifshitz spacetimes carrying fluxes of a true Maxwell field rather than a Proca field.  We find this can be done in five dimensions, where the source is provided by some Yang-Mills instanton or instanton-like configurations, through the equation
\be
d{*F}\sim F^a\wedge F^a\,.\label{formlang}
\ee
Indeed, the constant-time slice of the Lifshitz spacetime in five dimensions is a hyperbolic four-space in which one can construct $SU(2)$ Yang-Mills instanton or instanton-like configurations.  This observation allows us to construct Lifshitz spacetimes that carry electric charges of a true Maxwell field, provided that the spacetimes are also colored.  These provide first examples of electrically-charged Lifshitz spacetimes.

We carry out this construction in section 2.  We consider Einstein gravity coupled to a cosmological constant and $SU(2)$ Yang-Mills fields and a Maxwell field.  The key construction is the supergravity inspired $FFA$-type term that yields the right-hand side of equation (\ref{formlang}).  We also study the Green's functions for the dual operators associated with a minimally coupled $SU(2)$ scalar triplet and a $U(1)$ charged scalar.

In condensed matter theory, the low energy physics is more commonly described by Lifshitz theory with hyperscaling violation which breaks both Lorentz and scaling invariances of the theory. The gravitational dual can be constructed by introducing a dilaton field such that the spacetime is Lifshitz up to an overall dilaton scaling factor that violates the scaling symmetry \cite{Charmousis:2010zz,Gouteraux:2011ce,Iizuka:2011hg,Huijse:2011ef}.
Many Lifshitz-like solutions with hyperscaling violations have been found in literature [17-30]. In section 3, we demonstrate that by introducing a dilaton,  the hyperscaling violation is also possible in our system. In section 4, we present some exact black hole solutions that are asymptotic to the Lifshitz spacetimes with or without hyperscaling violations.  We present similar four-dimensional solutions in section 5.  The paper is concluded in section 6.

\section{Electrically charged Lifshitz spacetimes}

\subsection{The construction}

In this section, we consider $D=5$ gravity coupled to a cosmological constant $\Lambda$, $SU(2)$ Yang-Mills fields $A^{a}_\mu$ ($a=1,2,3$) and Maxwell field $A_\mu$. The Lagrangian is given by
\be
{\cal L} = \sqrt{-g} \Big( R- 3\Lambda - \fft1{4g_s^2} F^a_{\mu\nu} F^{a\mu\nu} - \fft14 F_{\mu\nu}F^{\mu\nu}\Big) - \fft{\sigma}{2g_s^2}\varepsilon^{\mu_1\mu_2\mu_3\mu_4\rho} F^a_{\mu_1\mu_2}
F^a_{\mu_3\mu_4} A_\rho\,,\label{d5lag1}
\ee
where $\varepsilon$ is a totally-antisymmetric tensor density, defined by $\varepsilon^{01234}=1$.  The Yang-Mills and Maxwell field strengths are
\be
F^a_{\mu\nu}=\partial_\mu A^a_\nu-\partial_\nu A^a_\mu+\epsilon^{abc}A^b_\mu A^c_\nu\,,\qquad
F_{\mu\nu}=\partial_\mu A_\nu-\partial_\nu A_\mu\,.
\ee
The most general two-derivative Lagrangian that respects all the symmetries will include also an $FFA$ term for the Maxwell field alone.  We shall not consider this term here since it has no effect on the solutions that we shall construct.

It should be emphasized that all the fields we consider above are of garden variety and have been observed in nature. The topological $FFA$-term is inspired by five-dimensional supergravities and it is unique in five dimensions.  The Lagrangian (\ref{d5lag1}) cannot be generalized to other dimensions with our content of fields.  The theory is specified by three non-trivial coupling constants, $(\Lambda, g_s, \sigma)$ and we let these coupling constants be arbitrary to begin with. The equations of motion are
\bea
D_\mu F^{a\mu\nu} &\equiv &\nabla_\mu F^{a\mu\nu}+\epsilon^{abc}A_\mu^b F^{c\mu\nu}=\sigma
\epsilon^{\nu\mu_1\mu_2\mu_3\mu_4} F^a_{\mu_1\mu_2} F_{\mu_3\mu_4}\,,\cr
\nabla_{\mu} F^{\mu\nu} &=&\fft{\sigma}{2g_s^2} \epsilon^{\nu\mu_1\mu_2\mu_3\mu_4} F^a_{\mu_1\mu_2} F^a_{\mu_3\mu_4}
\,,\cr
R_{\mu\nu} &=& \Lambda g_{\mu\nu} + \fft1{2g_s^2} \big(g^{\rho\sigma}F^a_{\mu\rho}F^a_{\nu\sigma}-\ft1{6} F^2 g_{\mu\nu}\big)+\ft1{2} \big(g^{\rho\sigma}F_{\mu\rho}F_{\nu\sigma}-\ft1{6} F^2 g_{\mu\nu}\big)\,,\label{eom1}
\eea
where $\epsilon$ is a tensor, given by $\epsilon^{\mu_1\cdots\mu_5} = \varepsilon^{\mu_1\cdots\mu_5}/\sqrt{-g}$.  Note that the second equation above becomes of (\ref{formlang}) in the form language.

The solutions to (\ref{eom1}) can be be embedded in $D=5$, ${\cal N}=4$, $SU(2)\times U(1)$ gauged supergravity provided that
\be
\Lambda=-2g_s^2\,,\qquad \sigma=\pm \ft14\,,\qquad
(F^a)^2=2F^2\,.\label{susycond}
\ee
(The bosonic sector of $D=5$, ${\cal N}=4$, $SU(2)\times U(1)$ gauged supergravity \cite{Romans:1985ps} contain in addition a scalar and rank-2 antisymmetric tensor.)

As we have discussed in the introduction, the motivation for us to consider the theory (\ref{d5lag1}) is to construct Lifshitz spacetimes that carry Maxwell fluxes.  For a charged solution that preserves the Lifshitz symmetry, the Ansatz for the Maxwell field must be
\be
A=q r^z dt\,,
\ee
where $q$ is a constant.  It is easy to verify that it will not satisfy the Maxwell equation of motion $d{*F}=0$, since ${*F}=z q r^3 dx_1\wedge dx_2\wedge dx_3$.  This problem is circumvented for a Proca field where ${*A}$ can provide a source with $d{*F}\sim {*A}$.  In our theory (\ref{d5lag1}), the source is provided by the Yang-Mills instanton(-like) configurations, via the second equation in (\ref{eom1}).  To see this explicitly, we consider the Ansatz for the $SU(2)$ Yang-Mills fields that respect the Lifshitz symmetry, given by
\be
A^a=p r\, dx_a\qquad \longrightarrow\qquad  F^a\wedge F^a = 3p^3\, \Omega_\4\,.
\ee
where $p$ is a constant and  $\Omega_\4$ is the volume form of the four-dimensional hyperbolic space
\be
ds_4^2 = \fft{dr^2}{r^2} + r^2 (dx_1^2 + dx_2^2 + dx_3^2)
\ee
that is a constant-time slice of the Lifshitz spacetime. It is easy to verify that when $p=1$, we have
\be
F^a={*}_4F^a\,,\qquad D_i F^{aij}=0\,.
\ee
Thus our $p=1$ Ansatz gives rise precisely to an Yang-Mills instanton in the hyperbolic 4-space.  We refer the cases of general $p$ to instanton-like configurations.

The second equation in (\ref{eom1}) implies that
\be
q=-\fft{4\sigma}{g_s^2} p^3\,.
\ee
In other words, the electric flux is provided by Yang-Mills instanton(-like) numbers. It is straightforward to construct the full Lifshitz solution, given by
\bea
ds^2 &=& - r^{2z} dt^2 + \fft{dr^2}{r^2} + r^2 (dx_1^2 + dx_2^2 + dx_3^2)\,,\cr
A^a &=& p r\, dx_a\,,\qquad
A=q r^z dt\,,\qquad
q = \ft1{z}\sqrt{2(z-1)(z-2p^2+2)}\,,\cr
\sigma &=& -\fft{z q}{8p(z-1)}\,,\qquad
g_s^2 = \fft{p^2}{2(z-1)}\,,\qquad
\Lambda=-\ft13 (z^2 + 4z+ 7 + p^2(z-1))\,.\label{genlifs}
\eea
When $\sigma=0$, we have $q=0$ and the solution reduces to colored Lifshitz spacetimes obtained in \cite{Fan:2015yza}, with $p^2=\ft12(z+2)$.  In this $\sigma=0$ case, a Yang-Mills instanton, corresponding to $p=1$, cannot support a Lifshitz spacetime alone since it requires that $z=0$, and hence $g_s^2<0$.

   It is of interest to examine what happens when the Yang-Mills configuration is indeed an instanton $(p=1)$.  The solution becomes
\bea
ds^2 &=& - r^{2z} dt^2 + \fft{dr^2}{r^2} + r^2 (dx_1^2 + dx_2^2 + dx_3^2)\,,
\quad A^a = r\, dx_a\,,\quad
A=\sqrt{\ft{2(z-1)}{z}}\, r^z dt\,,\cr
g_s &=& \sqrt{\ft{1}{2(z-1)}}\,,\qquad
\sigma=-\sqrt{\ft{z}{32(z-1)}}\,,\qquad
\Lambda=-\ft13(z+2)(z+3)\,.
\eea
The general Lifshitz spacetimes (\ref{genlifs}) are specified by two parameters $(z,p)$, which implies that the three coupling parameters $(\Lambda, g_s, \sigma)$ of the theory are not independent.  It turns out that the three conditions (\ref{susycond}) cannot be all satisfied for our Lifshitz spacetimes and hence the solutions cannot be embedded in $SU(2)\times U(1)$ gauged supergravity.

\subsection{Green's functions in boundary field theories}

Having obtained the electrically-charged and colored Lifshitz spacetimes, we would like to study the implications on the boundary field theories.  We first consider a minimally-coupled $U(1)$-charged scalar
\be
\mathcal{L}_{\phi}=\sqrt{-g}(-D_\mu \phi D^{\mu*}\phi^*-m^2\phi \phi^*)\,,
\ee
where $D_\mu=\partial_\mu-i A_\mu$.  The covariant equation of motion is:
\be
\frac{1}{\sqrt{-g}}D_\mu(\sqrt{-g}g^{\mu\nu}D_\nu \phi)-m^2\phi=0\,.
\ee
Taking the Fourier transformation
\be
\phi(r,t,x)=\varphi(r)e^{-{\rm i}\omega t+{\rm i} k x_1}\,,
\ee
we can derive a more explicit equation for $\varphi(r)$
\be
\varphi''+\frac{z+4}{r}\varphi'+\Big(\frac{(\omega+A_t)^2}{r^{2z+2}}-
\frac{k^2}{r^4}-\frac{m^2}{r^2}\Big)\varphi=0\,,
\ee
where the momentum has been chosen along the $x_1$ direction due to the rotational invariance. In the asymptotic $r\rightarrow \infty$ limit, we have:
\be
\varphi\rightarrow \frac{A}{r^{\Delta_{-}}}+\frac{B}{r^{\Delta_{+}}},\qquad \Delta_\pm=\ft12(z+3)\pm \mu\,,
\ee
where $A,B$ are integration constants and $\mu=\frac 12\sqrt{(z+3)^2+4m^2-4q^2}$.  For the dual operators to have real dimensions, we need imposing the Breitenlohner-Freedman type of bound, namely
\be
m^2\geq m^2_{BF},\qquad m^2_{BF}=-\frac{(z+3)^2}{4}+q^2\,.
\ee
We shall work in standard quantization, where the vev of the dual operator is defined by $\langle O_+ \rangle=B$ and the two-point function can be derived as: $G_{O_+}=B/A$.\footnote{For a certain mass range $m^2_{BF}\leq m^2\leq m^2_{BF}+1$, the $A$ mode is also normalizable and one can also work in the alternative quantization, where the vev of the dual operator is now defined by $\langle O_- \rangle=A$ and the two-point function is given by $G_{O_-}=A/B$.}

For vanishing $\omega$, Eq.~(23) can be analytically solved in terms of Bessel functions:
\be
\varphi=r^{-\frac{z+3}{2}} K_\mu(k/r)\,.
\ee
In the boundary limit, we find
\be
\varphi \rightarrow \fft{k^{-\mu}\, \Gamma(\mu)}{2^{1-\mu} r^{\Delta_-}}+
\fft{k^\mu\, \Gamma(-\mu)}{2^{1+\mu} r^{\Delta_+}}\,.
\ee
The Green's function for $\omega=0$ can be readily read off:
\be
G_{O_+}(\omega=0,k)=(\ft12k)^{2\mu} \fft{\Gamma(-\mu)}{\Gamma(\mu)}\,.
\ee
For non-vanishing $\omega$, Eq.(23) cannot be analytically solved except for the relativistic case $z=1$.

We now consider a minimally-coupled $SU(2)$ scalar triplet
\be
{\cal L}_{\phi^a}=\sqrt{-g}\big(-\ft12 D_\mu \phi^a D^{\mu}\phi^a-\ft12 m^2 \phi^a\phi^a\big)\,,
\ee
where $D_\mu \phi^a=\partial_\mu \phi^a+\epsilon^{abc}A^b_\mu \phi^c$.  The covariant equations are
\be
\frac{1}{\sqrt{-g}}D_\mu(\sqrt{-g}g^{\mu\nu}D_\nu \phi^a)-m^2\phi^a=0\,.
\ee
The wave equations were analysed in \cite{Fan:2015yza} in the momentum space and they can be solved for zero frequency ($\omega=0$) or zero momenta ($k^2=k^ik^i=0$), giving rise to the Green's functions
\be
G_{O_+}(\omega=0,k)=(\ft12k)^{2\nu} \fft{\Gamma(-\nu)}{\Gamma(\nu)}\,,\qquad
G_{O_+}(\omega,k=0)=\Big(\fft{{\rm i}\omega}{2z}\Big)^{2\nu} \fft{\Gamma(-\fft{\nu}{z})}{\Gamma(\fft{\nu}{z})}\,,
\ee
where
\be
\nu=\ft12\sqrt{(z+3)^2+4(m^2 + 2p^2)}\,.
\ee

It is interesting note that the Yang-Mills fluxes, which are magnetic, contributes a positive constant shift to the effective mass-square of the scalar triplet.  On the other hand, the Maxwell flux, which is electric, contribute an negative constant shift.  It follows that
the condensed matter system becomes unstable for large $q$, but remains stable for large $p$.  For the $SU(2)$ scalars that are also charged, the two characteristic powers of falloffs at large $r$ are given by
\be
\Delta_\pm =\ft12(z+3) \pm \ft12 \sqrt{(z+3)^2 + 4m^2 + 8p^2 - 4 q^2}\,.
\ee

\section{Dilaton and hyperscaling violation}

As discussed in the introduction, condensed matter systems are more commonly described by the Lifshitz theory with some hyperscaling violation. In the gravity side, this can be achieved by an $r$-dependent scalar field.  It follows that Lifshitz-like solutions with hyperscaling violations carrying electric fluxes are easier to come by than those electrically-charged Lifshitz solutions. (See, e.g.~the references listed in the relevant part of the Introduction.) In this section, we introduce a dilaton $\Phi$ to (\ref{d5lag1}) and propose a new Lagrangian
\bea
{\cal L}_{\rm \sst{YM}}&=&\sqrt{-g}\Big[ e^{-2\Phi}\Big( R- 3\Lambda +\ft12\omega (\partial\Phi)^2 - \fft1{4g_s^2} F^a_{\mu\nu} F^{a\mu\nu}\Big) - \fft14 e^{2\Phi} F_{\mu\nu}F^{\mu\nu}\Big]\cr
&&- \fft{\sigma}{2g_s^2}\varepsilon^{\mu_1\mu_2\mu_3\mu_4\rho} F^a_{\mu_1\mu_2}
F^a_{\mu_3\mu_4} A_\rho\,.\label{d5lag2}
\eea
The subscript of the Lagrangian denotes that it is written in the ``Yang-Mills frame'', in which the Yang-Mills fields do not ``see'' the dilaton $\Phi$ explicitly.  We introduce a new coupling constant $\omega$ for the dilaton.  The effective Yang-Mills coupling in this frame is then given by
\be
g^{\rm eff}_{\rm YM} = g_s\,\langle e^\Phi\rangle\,.
\ee
It is instructive to write the Lagrangian also in the Einstein frame, which we find to be
\bea
{\cal L}_{\rm \sst{Ein}}&=&\sqrt{-g}\Big( R- 3\Lambda e^{-\lambda\phi} -\ft12 (\partial\phi)^2 - \fft1{4g_s^2} e^{\lambda\phi} F^a_{\mu\nu} F^{a\mu\nu} - \fft14 e^{-2\lambda\phi} F_{\mu\nu}F^{\mu\nu}\Big)\cr
&&- \fft{\sigma}{2g_s^2}\varepsilon^{\mu_1\mu_2\mu_3\mu_4\rho} F^a_{\mu_1\mu_2}
F^a_{\mu_3\mu_4} A_\rho\,,\label{d5lag3}
\eea
where $\lambda$ and $\phi$ are related to $\omega$ and $\Phi$ as follows
\be
\lambda^2 = \fft{48}{9(32-3\omega)}\,,\qquad \phi=-\fft{4}{3\lambda} \Phi\,.
\ee
The metrics in the Einstein and Yang-Mills frames are related by
\be
ds_{\rm \sst{Ein}}^2 = e^{\lambda\phi} ds_{\rm \sst{YM}}^2\,.
\ee
Note that when $\omega >32/3$, the dilaton coupling constant $\lambda$ becomes purely imaginary.  The reality of the Lagrangian (\ref{d5lag3}) can be restored by letting $\phi\rightarrow {\rm i}\phi$.  This implies that the dilaton $\phi$ for $\omega >32/3$ has wrong kinetic sign and hence ghost-like.  This observation is not apparent in the Lagrangian (\ref{d5lag2}).

We obtain a class of colored and charged Lifshitz solutions with hyperscaling violations.  In the Yang-Mills frame, the solutions are
\bea
ds^2_{\rm \sst{YM}} &=& - r^{2z} dt^2 + \fft{dr^2}{r^2} + r^2 (dx_1^2 + dx_2^2 + dx_3^2)\,,\cr
A^a &=& p r\, dx_a\,,\qquad A=q r^{z + \fft32\theta} dt\,,\qquad
\Phi=-\ft34\theta \log r\,,\qquad
\omega = 8 - \fft{16z}{3\theta}\,,\cr
\sigma &=& -\ft18\sqrt{\fft{2z+3\theta +4-4p^2}{(z-1)p^2}}\,,\qquad
q=\fft{2\sqrt{(z-1)(2z+3\theta+4-4p^2)}}{2z+3\theta}\,,\cr
\Lambda &=& -\ft13\Big(\ft94 \theta^2 + \ft32 (2z+5)\theta + z^2 + 4z+6 + p^2(z-1)\Big)\,.
\eea
They are specified by three parameters, $(z,p,\theta)$, which are determined by $(\Lambda, g_s, \sigma,\omega)$ of the theory.  In the Einstein frame, the metric is simply $ds_{\rm Ein}^2 = r^\theta ds_{\rm YM}^2$, with
\be
\lambda^2 = \fft{2\theta}{3(2z+\theta)}\,,\qquad \phi=\fft{\theta}{\lambda} \log r\,.
\ee
When $\lambda=0$, corresponding to $\omega\rightarrow -\infty$, we have $\theta=0$, and the dilaton decouples from the theory.  The solution reduces to the one obtained in section 2.1.

It should be commented that although the metric in the Yang-Mills frame is Lifshitz, the Lifshitz symmetry is still broken in the full theory since the dilaton $\Phi$ is $r$-dependent. Of course, any field, such as graviton and Yang-Mills fields, who does not see the dilaton is not sensitive to the hyperscaling violation.  On the other hand, the Maxwell field is indeed modified by the hyperscaling violation.

   A necessary condition for the theory to be embedded into supergravity is
$\lambda^2=1$, which implies $z=0$ or $\theta\rightarrow \infty$.  Thus unfortunately, our solutions cannot be embedded in five-dimensional gauged
supergravities.

\section{Exact Lifshitz black holes}

Having obtained the charged and colored Lifshitz spacetimes, it is of interest to construct black hole solutions.  Unfortunately, we could not find any such exact solutions within the theories considered in the previous sections.  By introducing a new Maxwell field ${\cal F}=d{\cal A}$, we construct a class of exact Lifshitz black holes that are charged under ${\cal A}$.  The Lagrangian for ${\cal A}$ in the Yang-Mills frame is given by
\be
{\cal L}_{\rm \sst{YM}}^{\cal A} = -\ft14 \sqrt{-g} e^{-2\Phi} {\cal F}_{\mu\nu}{\cal F}^{\mu\nu} \,.
\ee
We find a class of exact black hole solutions
\bea
ds^2 &=& -r^{2z} \tilde f dt^2 + \fft{dr^2}{r^2\tilde f} + r^2 (dx_1^2 + dx_2^2 + dx_3^2)\,,\qquad \tilde f = 1- \fft{ Q^2}{2(z-1)r^{2(z-1)}}\,,\cr
A^a &=& p r\, dx_a\,,\qquad A=q r^{2(z-2)} dt\,,\qquad {\cal A}=Q r dt
\qquad \Phi=-\ft12(z-4) \log r\,,\cr
\sigma &=& - \fft{\sqrt{z-1-p^2}}{4p\sqrt{z-1}}\,,\qquad q=\fft{\sqrt{(z-1)(z-1-p^2)}}{z-2}\,,\cr
\Lambda &=& -\ft13 (z-1)(4z-3+p^2)\,,\qquad \omega = \fft{32}{z-4}\,.
\eea
In other words, the hyperscaling violation parameter of the the corresponding asymptotic Lifshitz spacetime is constrained to be $\theta=\fft23 (z-4)$.  In the Einstein frame, the dilaton coupling constant is given by
\be
\lambda^2 = \fft{z-4}{6(z-1)}\,.
\ee
Thus for the dilaton to be non-ghost like, we must have $z\ge 4$.  When $z=4$, the dilaton decouples and the corresponding Lifshitz vacuum has no hyperscaling violation.

   All the black holes have vanishing mass, but satisfy
\be
T dS + \Phi_e dQ_e=0\,,\qquad TS + \Phi_e Q_e=0\,,\label{fl}
\ee
with
\be
T=\ft{z-1}{2\pi} r_0^z\,,\qquad S=\ft14 r_0^{z-1}\,,\qquad
\Phi_e=-r_0 Q\,,\qquad Q_e=\ft{1}{16\pi}Q\,,
\ee
where $r_0$ is the horizon and $\tilde f(r_0)=0$.  There have been many examples of such Lifshitz black holes in literature \cite{Zingg:2011cw,Liu:2014dva,Fan:2014ixa,Fan:2014ala,
Fan:2015yza,Bravo-Gaete:2015xea}.

\section{$D=4$ Lifshitz spacetimes with or without hyperscaling violations}

We now consider the analogous construction in four dimensions.  The matter field content consists of the $SU(2)$ Yang-Mills fields, a Maxwell field and a dilaton $\Phi$.
The Lagrangian in the Yang-Mills frame is given by
\be
{\cal L}_{\rm \sst{YM}} = \sqrt{-g} e^{2\Phi} \Big(R -2\Lambda + \ft12 \omega (\partial\Phi)^2 -
\fft{1}{4g_s^2} F^a_{\mu\nu} F^{a\mu\nu} -\fft14 {\cal F}^2\Big)\,.
\ee
It can be converted to the Einstein frame using the conformal scaling
\be
ds^2_{\rm \sst{Ein}} = e^{\lambda\phi} ds_{\rm \sst{YM}}^2\,,\qquad
\hbox{with}\qquad \lambda^2 =\fft{4}{12-\omega}\,,\qquad \phi=-\fft{2}{\lambda}\Phi\,.
\ee
The Lagrangian in the Einstein frame is then given by
\be
{\cal L}_{\rm \sst{Ein}} = \sqrt{-g}\Big(R -2\Lambda e^{-\lambda\phi} - \ft12 (\partial\phi)^2 -
\fft{1}{4g_s^2} e^{\lambda\phi} F^a_{\mu\nu} F^{a\mu\nu} -\fft14 e^{\lambda\phi} {\cal F}^2\Big)\label{d4lag2}\,.
\ee
Turning off the vector field ${\cal A}$, we obtain a class of colored Lifshitz vacua with hyperscaling violations.  In the Einstein frame, they are given by
\bea
ds &=& r^{\theta}\Big(-r^{2z} dt^2 + \fft{dr^2}{r^2} + r^2 (dx_1^2 + dx_2^2)\Big)\,,\cr
\phi &=& \fft{\theta}{\lambda} \log r\,,\qquad A^3=0\,,\qquad
A^a = pr dx_a\,,\qquad a=1,2,\cr
\lambda&=&\sqrt{\fft{\theta}{2z+\theta}}\,,\qquad
g_s^2 = \fft{z+1+\theta}{2(z-1)}\,,\qquad p=\sqrt{1 + z + \theta}\,,\cr
\Lambda &=& -\ft12(\theta^2 + (2z+3)\theta + z^2+2z+3)\,.
\eea
When $\lambda=0$, we have $\theta=0$ and consequently the dilaton decouples.  The resulting Lifshitz vacua with no hyperscaling violation was obtained previously in \cite{Devecioglu:2014iia,Fan:2015yza}.  The Lagrangian (\ref{d4lag2}) also admits an exact black hole solution charged under the Maxwell field ${\cal A}$ for $\theta = z-3$.  It is given by
\bea
ds &=& r^{z-3}\Big(-r^{2z} \tilde fdt^2 + \fft{dr^2}{r^2\tilde f} + r^2 (dx_1^2 + dx_2^2)\Big)\,,\qquad \tilde f=1-\fft{Q^2}{2(z-1)r^{2(z-1)}}\cr
{\cal A} &=& Q r dt\,,\quad (A^1,A^2,A^3)= \sqrt{2(z-1)}\, r (dx_1,dx_2,0)\,,
\quad \phi=\sqrt{3(z-1)(z-3)}\, \log r\,,\cr
\lambda &=&\sqrt{\ft{z-3}{3(z-1)}}\,,\qquad
g_s^2 = 1\,,\qquad \Lambda = -\ft12(4z^2-7z+3)\,.
\eea
The solution with $\theta=0$, i.e.~$z=3$ was obtained in \cite{Fan:2015yza}.
It can be easily checked that the electrically charged black holes have zero mass, with the thermodynamical quantities $(T,S,\Phi_e, Q_e)$ satisfying (\ref{fl}).

\section{Conclusions}

In this paper we construct a class of Lifshitz spacetimes in five dimensions that carry electric fluxes of a Maxwell field.  The theory that makes this possible is Einstein gravity coupled to a cosmological constant, $SU(2)$ Yang-Mills fields as well as the Maxwell field.  The key construction is the supergravity-inspired $FFA$-type Chern-Simons term.  This term provides the Maxwell equation with a Yang-Mills instanton(-like) source.  The allowed dynamic exponents of the Lifshitz solutions are $z>1$.  We studied the scalar wave equations of the electrically-charged and/or $SU(2)$-colored scalars and found that the electric background added a negative constant shift to the effective scalar mass-square whilst the magnetic Yang-Mills background added a positive constant shift.

    We introduced a dilaton to the system and obtained a class of Lifshitz
spacetimes with hyperscaling violations for $z>1$ with arbitrary parameter $\theta$.  We also constructed some exact black hole solutions with vanishing mass, but satisfying a non-trivial thermodynamical first law (\ref{fl}).  We constructed analogous solutions with or without hyperscaling violations in four dimensions.

   The matter fields in our construction are all of garden variety, namely
the Maxwell and Yang-Mills fields rather than the Proca field or more exotic ones. Our Lisfhitz spacetimes carrying electric fluxes appear to be first in literature and should provide interesting gravity duals for studying some strongly couple condensed matter systems.

\section*{Acknowledgement}

We are grateful to Sera Cremonini and Chris Pope for useful discussions. H.L.~is grateful to the Mitchell Institute at Taxas A\&M University for hospitality during the late stage of this work. Z.-Y. Fan is supported in part by NSFC Grants NO.10975016, NO.11235003 and NCET-12-
0054; The work of H.L. is supported in part by NSFC grants NO.1
1175269, NO.11475024 and NO.11235003.

\end{document}